\documentclass[runningheads]{llncs}
\usepackage{multirow}
\usepackage{graphicx}
\usepackage{color}
\usepackage[misc]{ifsym} 
\begin{document}
%
\title{Re-thinking and Re-labeling LIDC-IDRI for Robust Pulmonary Cancer Prediction}
%
%
\author{Hanxiao Zhang\inst{1} \and
Xiao Gu\inst{2} \and
Minghui Zhang\inst{1} \and
Weihao Yu\inst{1} \and
Liang Chen\inst{3} \and
Zhexin Wang\inst{3} \and
Feng Yao\inst{3} \and
Yun Gu\inst{1,4}\textsuperscript{(\Letter)} \and
Guang-Zhong Yang\inst{1}}
\authorrunning{H. Zhang et al.}

%

\institute{Institute of Medical Robotics, Shanghai Jiao Tong University, Shanghai, China
\email{geron762@sjtu.edu.cn}\\
\and
Imperial College London, London, UK
\and
Department of Thoracic Surgery, Shanghai Chest Hospital, Shanghai Jiao Tong University, Shanghai, China
\and
Shanghai Center for Brain Science and Brain-Inspired Technology, Shanghai, China
}

\maketitle              
\begin{abstract}
The LIDC-IDRI database is the most popular benchmark for lung cancer prediction. However, with subjective assessment from radiologists, nodules in LIDC may have entirely different malignancy annotations from the pathological ground truth, introducing label assignment errors and subsequent supervision bias during training. 
The LIDC database thus requires more objective labels for learning-based cancer prediction.
Based on an extra small dataset containing 180 nodules diagnosed by pathological examination, we propose to re-label LIDC data to mitigate the effect of original annotation bias verified on this robust benchmark.
We demonstrate in this paper that providing new labels by similar nodule retrieval based on metric learning would be an effective re-labeling strategy.
Training on these re-labeled LIDC nodules leads to improved model performance, which is enhanced when new labels of uncertain nodules are added. 
We further infer that re-labeling LIDC is current an expedient way for robust lung cancer prediction while building a large pathological-proven nodule database provides the long-term solution.

\end{abstract}
\section{Introduction}

The LIDC-IDRI (Lung Image Database Consortium and Image Database Resource Initiative) \cite{armato2011lung} is a leading source of public datasets. Since the introduction of LIDC, it is used extensively for lung nodule detection and cancer prediction using learning-based methods \cite{han2015texture,shen2015multi,shen2016learning,shen2017multi,hussein2017risk,wu2018joint,xie2018knowledge,liu2019multi,liao2021learning}.

When searching papers in PubMed\footnote{https://pubmed.ncbi.nlm.nih.gov/} with the following filter:{\color{blue}{\textit{ ("deep learning" OR convolutional) AND (CT OR "computed tomography") AND (lung OR pulmonary) AND (nodule OR cancer OR "nodule malignancy") AND (prediction OR classification)}}}, among 53 papers assessed for eligibility of nodule malignancy classification, 40 papers used LIDC database, 5 papers used NSLT (National Lung Screening Trial) database \footnote{\url{https://cdas.cancer.gov/datasets/nlst/}} \cite{national2011national,national2011reduced,kramer2011lung} (no exact nodule location provided), and 8 papers used other individual datasets. LIDC is therefore the most popular benchmark in cancer prediction research.


A careful examination of the LIDC database, however, reveals several potential issues for cancer prediction. During the annotation of LIDC, characteristics of nodules were assessed by multiple radiologists, where the rating of malignancy scores (1 to 5) was based on the assumption of a \textbf{60-year-old male smoker}. Due to the lack of clinical information, these malignancy scores were subjective. Although a subset of LIDC cases possesses patient-based pathological diagnosis \cite{mcnitt2007lung}, its nodule-level binary labels can not be confirmed.

Since it is hard to recapture the pathological ground truth for each LIDC nodule, we apply the extra SCH-LND dataset~\cite{zhang2020learning} with pathological-proven labels, which is used not only for establishing a truthful and fair evaluation benchmark but also for transferring pathological knowledge for different clinical indications.


In this paper, we first assess the nodule prediction performances of LIDC driven model in six scenarios and their fine-tuning effects using SCH-LND with detailed experiments. Having identified the problems of the undecided binary label assignment scheme on the original LIDC database and unstable transfer learning outcomes, we seek to re-label LIDC nodule classes by interacting with the SCH-LND. The first re-labeling strategy adopts the state-of-the-art nodule classifier as an end-to-end annotator, but it has no contribution to LIDC re-labeling. The second strategy uses metric learning to learn similarity and discrimination between the nodule pairs, which is then used to elect new LIDC labels based on the similarity ranking in a pairwise manner between the under-labeled LIDC nodule and each nodule of SCH-LND. 
Experiments show that the models trained with re-labeled LIDC data created by metric learning model not only resolve the bias problem of the original data but also transcend the performance of our model, especially when the new labels of the uncertain subset are added. 
Further statistical results demonstrate that the re-labeled LIDC data suffers class imbalance problem, which indicates us to build a larger nodule database with pathological-proven labels.

\section{Materials}
\label{sec::dataset}
\noindent\textbf{LIDC-IDRI Database:}
According to the practice in \cite{setio2017validation}, we excluded CT scans with slice thickness larger than 3 mm and sampled nodules identified by at least three radiologists.
We only involve solid nodules in SCH-LND and LIDC databases because giving accurate labels for solid nodules is of great challenge.

\noindent\textbf{Extra Dataset:}
The extra dataset called SCH-LND \cite{zhang2020learning} consists of 180 solid nodules (90 benign/90 malignant) with exact spatial coordinates and radii. Each sample is very rare because all the nodules are confirmed and diagnosed by immediate pathological examination via biopsy with ethical approval.


To regulate variant CT formats, CT slice thickness is resampled to 1mm/pixel if it is larger than 1 mm/pixel, while the X and Y axes are fixed to 512$\times$512 pixels. Each pixel value is unified to the HU (Hounsfield Unit) value before nodule volume cropping.

\section{Study Design}
\label{sec::preliminary_work}

\begin{figure}[!hbt]
\centering
\includegraphics[width = 120 mm]{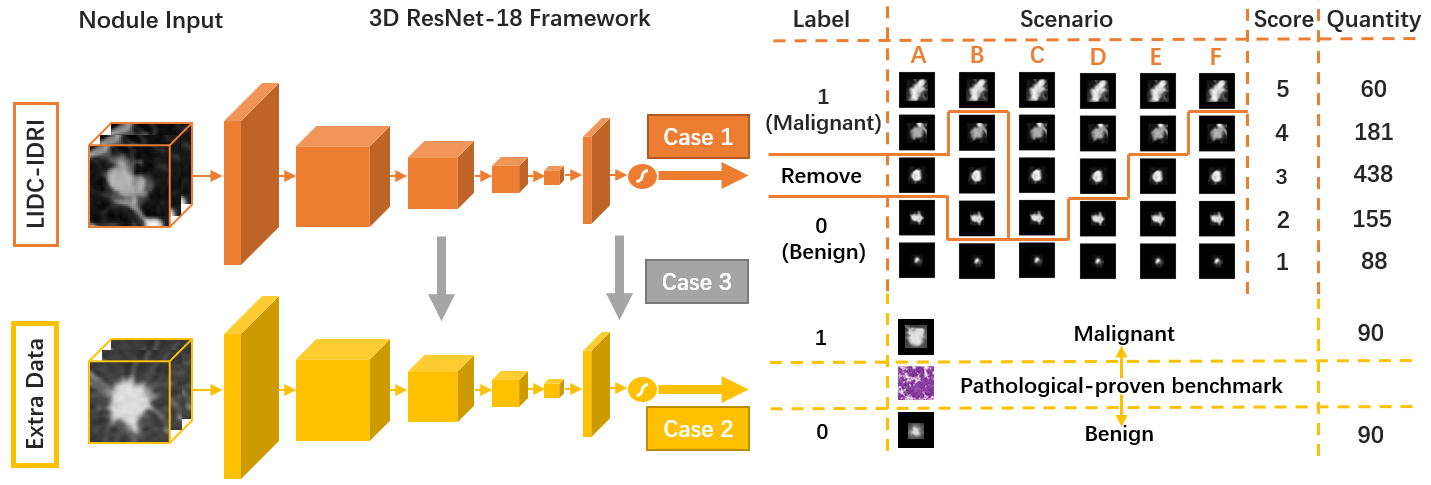}
\caption{Illustration of the study design for nodule cancer prediction. \textbf{Case 1:} training from scratch over the LIDC database after assigning nodule labels according to the average malignancy scores in 6 scenarios. \textbf{Case 2:} training over extra data based on accurate pathological-proven labels by 5-fold cross-validation. \textbf{Case 3:} testing or fine-tuning LIDC models of Case 1 using extra data.  
}
\label{fig1} 
\end{figure}

The preliminary study follows the instructions of Fig. \ref{fig1} where two types of cases (Case 1 and Case 2) conduct training and testing in each single data domain and one type of case (Case 3) involves domain interaction (cross-domain testing and transfer learning) between LIDC and SCH-LND. In Case 1 and Case 3, we identify 6 different scenarios by removing uncertain average scores (Scenarios A and B) or setting division threshold (Scenarios C, D, E, and F) to assign binary labels for LIDC data training. Training details are described in Section \ref{sec::Implementation}. 

To evaluate the model performance comprehensively, we additionally introduce Specificity (also called Recall$_{b}$, when treating benign as positive sample) and Precision$_{b}$ (Precision in benign class)  \cite{wu2019learning}, besides regular evaluation metrics including Sensitivity (Recall), Precision, Accuracy, and F1 score.

\begin{figure}[!hbt]
\centering
\includegraphics[width = 120 mm]{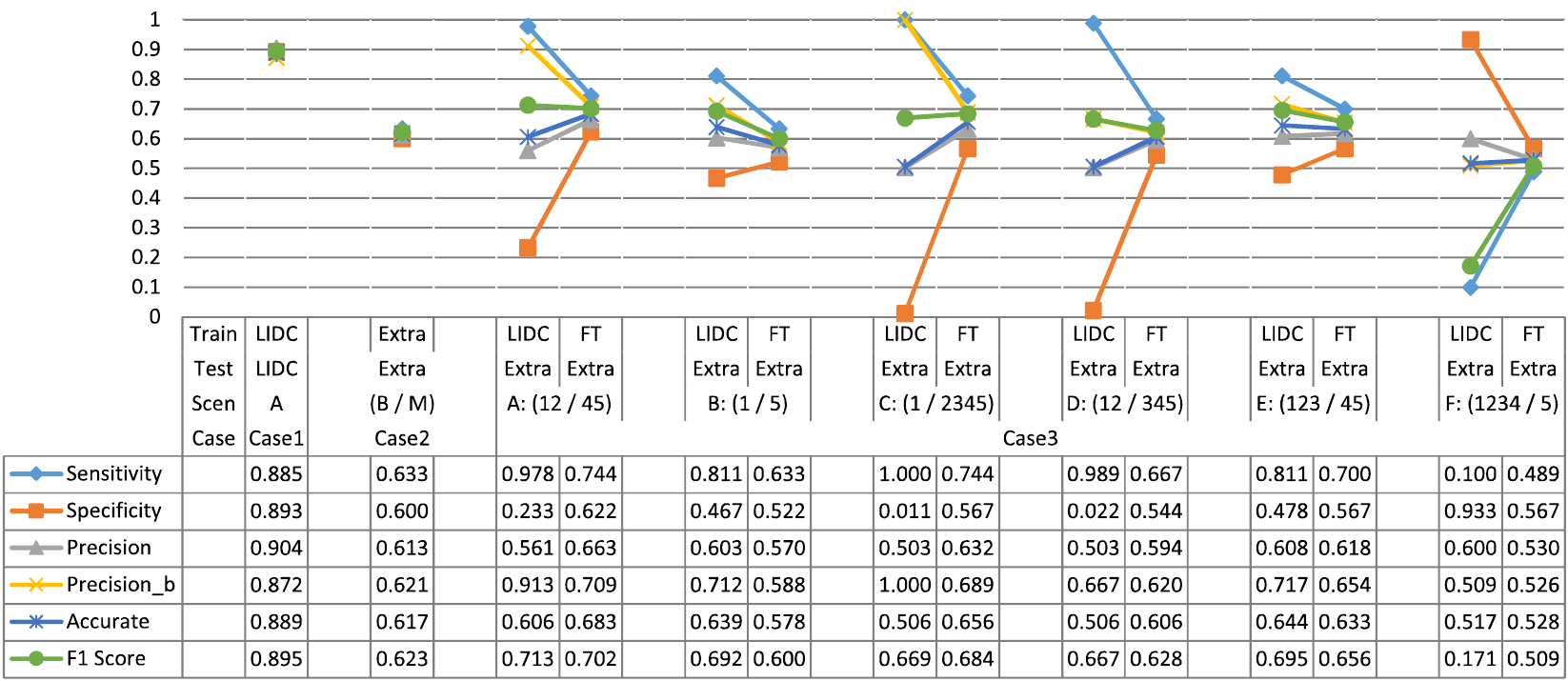}
\caption{Performance comparisons between different Cases or Scenarios (Scen) in Fig. \ref{fig1}. For instance, `A:(12/45)' represents `Scenario A' that treats LIDC scores 1 \& 2 as benign labels and scores 4 \& 5 as malignant labels. FT denotes fine-tuning using extra data by 5-fold cross-validation based on the pre-trained model in each scenario.
}
\label{fig2} 
\end{figure}

Based on the visual assessment of radiologists, human-defined nodule features can be easily extracted and classified by a commonly used model (3D ResNet-18 \cite{he2016deep}), whose performance can emulate the experts’ one (Fig. \ref{fig2}, Case 1). Many studies still put investigational efforts for better results across the LIDC board, overlooking inaccurate radiologists’ estimations and bad model capability in the real world.
However, once the same model is revalidated under the pathological-proven benchmark (Fig. \ref{fig2}, Case 3, Scenario A), its drawback is objectively revealed that LIDC model decisions take up too many false-positive predictions.
These two experimental outcomes raise a suspicion that whether the visual assessment of radiologists might have a bias toward malignant class.


To resolve this suspicion, we compare the performances of 6 scenarios in Case 3. Evidence reveals that, under the testing data from SCH-LND, the number of false-positive predictions has a declining trend when the division threshold moves from the benign side to the malignant side, but the bias problem is still serious when reaching Scenario E, much less of Scenario A and B.
Besides, as training on the SCH-LND dataset from scratch can hardly obtain a high capacity model (Fig. \ref{fig2}, Case 2), we use transfer learning in Case 3 to get the model fine-tuned on the basis of weights of different pre-trained LIDC models.

Observing the inter-comparison within each scenario in Case 3, transfer learning can push scattered metric values close. However, compared with Case 2, the fine-tuning technique would bring both positive and negative transfer, depending upon the property of the pre-trained model.

Thus, either for training from scratch or transfer learning process, the radiologists’ assessment of LIDC nodule malignancy can be hard to properly use. In addition to its inevitable assessment errors, there is a thorny problem to assign LIDC labels (how to set division threshold) and removing uncertain subset (waste of data). We thus expect to re-label the LIDC malignancy classes with the interaction of SCH-LND, to correct the assessment bias as well as utilize the uncertain nodules (average score=3). Two independent approaches are described in the following section.

\section{Methods}

We put forward two re-labeling strategies to obtain new ground truth labels on the LIDC database. The first strategy generates the malignancy label from a machine annotator: the state-of-the-art nodule classifier that has been pre-trained on LIDC data and fine-tuned on SCH-LND to predict nodule class. The second strategy ranks the top nodules’ labels using a machine comparator: a metric-based Network that measures the correlation between nodule pairs.

Considering that the knowledge from radiologists' assessments could be a useful resource, in each strategy, two modes of LIDC re-labeling are proposed. 
\textbf{For Mode 1 (Substitute):} LIDC completely accepts the re-label outcomes from other label machines. 
\textbf{For Mode 2 (Consensus):} The final LIDC re-label results would be decided by the consensus of label machine outcomes and its original label (Scenario A). In other words, this mode will leave behind the nodules with the same label and discard controversial ones, which may cause data reduction.
We evaluate the LIDC re-labeling effect by using SCH-LND to test the model which is trained with re-labeled data from scratch.

\subsection{Label Induction Using Machine Annotator}
The optimized model with fine-tuning technique can correct the learning bias initiated by LIDC data. Some fine-tuned models even surpass the LIDC model performance in large scales of evaluation metrics. We wonder whether the current best performance model can help classify and annotate new LIDC labels. Experiments will be conducted using two annotation models from Case 2 and Case 3 (Scenario A) in Section \ref{sec::preliminary_work}.

\subsection{Similar Nodule Retrieval Using Metric Learning}

\begin{figure}[!hbt]
\centering
\includegraphics[width = 120 mm]{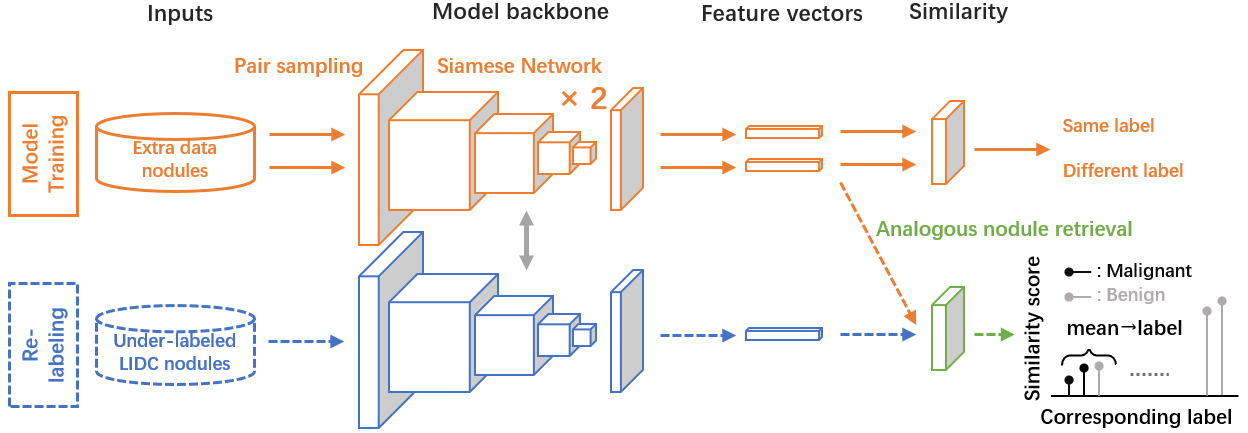}
\caption{The second strategy of LIDC re-labeling that using a metric learning model to search for the most similar nodules and give new labels.  
}
\label{fig3} 
\end{figure}

Metric learning \cite{bellet2015metric}\cite{kaya2019deep} provides a few-shot learning approach that aims to learn useful representations through distance comparisons. We use Siamese Network \cite{koch2015siamese}\cite{guo2017learning} in this study which consists of two networks whose parameters are tied to each other. Parameter tying guarantees that two similar nodules will be mapped by their respective networks to adjacent locations in feature space. 

For training a Siamese Network in Fig. \ref{fig3}, we pass the inputs in the set of pairs. Each pair is randomly chosen from SCH-LND and given the label whether two nodules of this pair are in the same class. Then these two nodule volumes are passed through the 3D ResNet-18 to generate a fixed-length feature vector individually. A reasonable hypothesis is given that: if the two nodules belong to the same class, their feature vectors should have a small distance metric; otherwise, their feature vectors will have a large distance metric. 
In order to distinguish between the same and different pairs of nodules when training, we apply contrastive loss over the Euclidean distance metric (similarity score) induced by the malignancy representation.

During re-labeling, we first pair each nodule from SCH-LND used in training up with an under-labeled LIDC nodule and sort each under-labeled nodule partner by their similarity scores. Then the new LIDC label is awarded by averaging the labels of the top 20$\%$ partner nodules in the ranking list of similarity scores.

\section{Experiments and Results}
\label{sec::experiments_and_results}

\subsection{Implementation}
\label{sec::Implementation}

We apply 3D ResNet-18 \cite{he2016deep} in this paper with adaptive average pooling (output size of 1$\times$1$\times$1) following the final convolution layer. For the general cancer prediction model, we use a fully connected layer and a Sigmoid function to output the prediction score (binary cross-entropy loss). While for Siamese Network, we use a fully connected layer to generate the feature vector (8 neurons). Due to various nodule sizes, the batch size is set to 1, and group normalization \cite{wu2018group} is adopted after each convolution layer.

All the experiments are implemented in PyTorch with a single NVIDIA GeForce GTX 1080 Ti GPU and learned using the Adam optimizer \cite{kingma2014adam} with the learning rate of 1e-3 (100 epochs) and that of 1e-4 for fine-tuning in transfer learning (50 epochs). The validation set occupies 20$\%$ of the training set in each experiment. All the experiments and results involving or having involved the training of SCH-LND are strictly conducted by 5-fold cross-validation.

\subsection{Quantitative Evaluation}
\label{sec::Evaluation}

\begin{table}[]
\centering
\caption{Performances of different re-labeling methods based on each mode of re-labeling strategies. Under-labeled LIDC data are chosen by their original average score.}
\label{tab:my-table}
\resizebox{\textwidth}{!}{%
\begin{tabular}{ccccccccccc}
\hline
\multicolumn{1}{c|}{Row} &
  \multicolumn{1}{c|}{\multirow{4}{*}{Baselines}} &
  \multicolumn{1}{c|}{Method} &
  \multicolumn{1}{c|}{Training} &
  \multicolumn{1}{c|}{Testing} &
  \multicolumn{1}{c|}{Sensitivity} &
  \multicolumn{1}{c|}{Specificity} &
  \multicolumn{1}{c|}{Precision} &
  \multicolumn{1}{c|}{Precision$_{b}$} &
  \multicolumn{1}{c|}{Accuracy} &
  F1 \\ \cline{1-1} \cline{3-11} 
\multicolumn{1}{c|}{1} &
  \multicolumn{1}{c|}{} &
  \multicolumn{1}{c|}{Case 3-A} &
  \multicolumn{1}{c|}{LIDC} &
  \multicolumn{1}{c|}{Extra} &
  \multicolumn{1}{c|}{0.9778} &
  \multicolumn{1}{c|}{0.2333} &
  \multicolumn{1}{c|}{0.5605} &
  \multicolumn{1}{c|}{0.9130} &
  \multicolumn{1}{c|}{0.6056} &
  0.7126 \\ \cline{1-1} \cline{3-5}
\multicolumn{1}{c|}{2} &
  \multicolumn{1}{c|}{} &
  \multicolumn{1}{c|}{Case 2} &
  \multicolumn{1}{c|}{Extra} &
  \multicolumn{1}{c|}{Extra} &
  \multicolumn{1}{c|}{0.6333} &
  \multicolumn{1}{c|}{0.6000} &
  \multicolumn{1}{c|}{0.6129} &
  \multicolumn{1}{c|}{0.6207} &
  \multicolumn{1}{c|}{0.6167} &
  0.6230 \\ \cline{1-1} \cline{3-5}
\multicolumn{1}{c|}{3} &
  \multicolumn{1}{c|}{} &
  \multicolumn{1}{c|}{Siamese} &
  \multicolumn{1}{c|}{Extra} &
  \multicolumn{1}{c|}{Extra} &
  \multicolumn{1}{c|}{0.6667} &
  \multicolumn{1}{c|}{0.6000} &
  \multicolumn{1}{c|}{0.6250} &
  \multicolumn{1}{c|}{0.6429} &
  \multicolumn{1}{c|}{0.6333} &
  0.6452 \\ \hline
 &
  \multicolumn{10}{c}{LIDC re-labeling} \\ \hline
\multicolumn{1}{c|}{} &
  \multicolumn{1}{c|}{Strategy} &
  \multicolumn{1}{c|}{Mode} &
  \multicolumn{1}{c|}{Method} &
  \multicolumn{1}{c|}{Under-label} &
  \multicolumn{1}{c|}{Sensitivity} &
  \multicolumn{1}{c|}{Specificity} &
  \multicolumn{1}{c|}{Precision} &
  \multicolumn{1}{c|}{Precision$_{b}$} &
  \multicolumn{1}{c|}{Accuracy} &
  F1 \\ \hline
\multicolumn{1}{c|}{4} &
  \multicolumn{1}{c|}{\multirow{4}{*}{Annotator}} &
  \multicolumn{1}{c|}{\multirow{2}{*}{Substitute}} &
  \multicolumn{1}{c|}{Case 2} &
  \multicolumn{1}{c|}{\multirow{4}{*}{1;2;4;5}} &
  \multicolumn{1}{c|}{0.5778} &
  \multicolumn{1}{c|}{0.5667} &
  \multicolumn{1}{c|}{0.5714} &
  \multicolumn{1}{c|}{0.5730} &
  \multicolumn{1}{c|}{0.5722} &
  0.5746 \\ \cline{1-1} \cline{4-4}
\multicolumn{1}{c|}{5} &
  \multicolumn{1}{c|}{} &
  \multicolumn{1}{c|}{} &
  \multicolumn{1}{c|}{Case 3-A} &
  \multicolumn{1}{c|}{} &
  \multicolumn{1}{c|}{0.4630} &
  \multicolumn{1}{c|}{0.6667} &
  \multicolumn{1}{c|}{0.5814} &
  \multicolumn{1}{c|}{0.5538} &
  \multicolumn{1}{c|}{0.5648} &
  0.5155 \\ \cline{1-1} \cline{3-4}
\multicolumn{1}{c|}{6} &
  \multicolumn{1}{c|}{} &
  \multicolumn{1}{c|}{\multirow{2}{*}{Consensus}} &
  \multicolumn{1}{c|}{Case 2} &
  \multicolumn{1}{c|}{} &
  \multicolumn{1}{c|}{0.8778} &
  \multicolumn{1}{c|}{0.3667} &
  \multicolumn{1}{c|}{0.5809} &
  \multicolumn{1}{c|}{0.7500} &
  \multicolumn{1}{c|}{0.6222} &
  0.6991 \\ \cline{1-1} \cline{4-4}
\multicolumn{1}{c|}{7} &
  \multicolumn{1}{c|}{} &
  \multicolumn{1}{c|}{} &
  \multicolumn{1}{c|}{Case 3-A} &
  \multicolumn{1}{c|}{} &
  \multicolumn{1}{c|}{0.8556} &
  \multicolumn{1}{c|}{0.3778} &
  \multicolumn{1}{c|}{0.5789} &
  \multicolumn{1}{c|}{0.7234} &
  \multicolumn{1}{c|}{0.6167} &
  0.6906 \\ \hline
\multicolumn{1}{c|}{8} &
  \multicolumn{1}{c|}{\multirow{4}{*}{Comparator}} &
  \multicolumn{1}{c|}{\multirow{2}{*}{Substitute}} &
  \multicolumn{1}{c|}{\multirow{4}{*}{Siamese}} &
  \multicolumn{1}{c|}{\textbf{1;2;4;5}} &
  \multicolumn{1}{c|}{0.6111} &
  \multicolumn{1}{c|}{0.6556} &
  \multicolumn{1}{c|}{0.6395} &
  \multicolumn{1}{c|}{0.6277} &
  \multicolumn{1}{c|}{0.6333} &
  0.6250 \\ \cline{1-1} \cline{5-5}
\multicolumn{1}{c|}{9} &
  \multicolumn{1}{c|}{} &
  \multicolumn{1}{c|}{} &
  \multicolumn{1}{c|}{} &
  \multicolumn{1}{c|}{\textbf{1;2;3;4;5}} &
  \multicolumn{1}{c|}{0.6778} &
  \multicolumn{1}{c|}{\textbf{0.6667}} &
  \multicolumn{1}{c|}{\textbf{0.6703}} &
  \multicolumn{1}{c|}{0.6742} &
  \multicolumn{1}{c|}{\textbf{0.6722}} &
  \textbf{0.6740} \\ \cline{1-1} \cline{3-3} \cline{5-5}
\multicolumn{1}{c|}{10} &
  \multicolumn{1}{c|}{} &
  \multicolumn{1}{c|}{\multirow{2}{*}{Consensus}} &
  \multicolumn{1}{c|}{} &
  \multicolumn{1}{c|}{1;2;4;5} &
  \multicolumn{1}{c|}{0.8000} &
  \multicolumn{1}{c|}{0.3778} &
  \multicolumn{1}{c|}{0.5625} &
  \multicolumn{1}{c|}{0.6538} &
  \multicolumn{1}{c|}{0.5889} &
  0.6606 \\ \cline{1-1} \cline{5-5}
\multicolumn{1}{c|}{11} &
  \multicolumn{1}{c|}{} &
  \multicolumn{1}{c|}{} &
  \multicolumn{1}{c|}{} &
  \multicolumn{1}{c|}{1;2;3;4;5} &
  \multicolumn{1}{c|}{0.7333} &
  \multicolumn{1}{c|}{0.5889} &
  \multicolumn{1}{c|}{0.6408} &
  \multicolumn{1}{c|}{0.6883} &
  \multicolumn{1}{c|}{0.6611} &
  0.6839 \\ \hline
\end{tabular}%
}
\end{table}

To evaluate the first strategy using machine annotator, we first use Case 2 model to re-label LIDC nodules (a form of 5-fold cross-validation) other than the uncertain subset (original average score = 3). The re-labeled nodules are then fed into the 3D ResNet-18 model, which will be trained from scratch and tested on the corresponding subset of SCH-LND for evaluation. 
The result ($4^{th}$ row) shows that although this action greatly fixes label bias to a balanced state, this group of new labels can hardly build a model tested well on SCH-LND. 
Contrary to common sense, the state-of-the-art nodule classifier makes re-label performance worse ($5^{th}$ row), which is much lower than that of learning from scratch using SCH-LND ($2^{nd}$ row), indicating that the best model optimized with fine-tuning technique is not suitable for LIDC re-labeling.
The initial two experiments adopting Mode 2 (Consensus) achieved better comprehensive outcomes than Mode 1 (Substitute) but with low Specificity.

Metric learning takes a different re-label strategy that retrieves similar nodules according to the distance metric. Metric learning on a small dataset can obtain a better performance ($3^{rd}$ row) compared with general learning from scratch ($2^{nd}$ row). The re-label outcomes ($8^{th}$ and $9^{th}$ row) also show great comprehensive improvement over baselines by Mode 1, where the re-labeling of uncertain nodules (average score=3) is an important contributing factor. 

Overall, there is a trade-off between Mode 1 and Mode 2. But Mode 2 seems to remain the LIDC bias property because testing results often have low Specificity and introduce data reduction.
Re-labeling by consensus (Mode 2) may integrate the defects of both original labels and models, especially for malignant labels, while re-labeling uncertain nodules can help mitigate the defect of Mode 2.

\begin{figure}[!hbt]
\centering
\includegraphics[width = 120 mm]{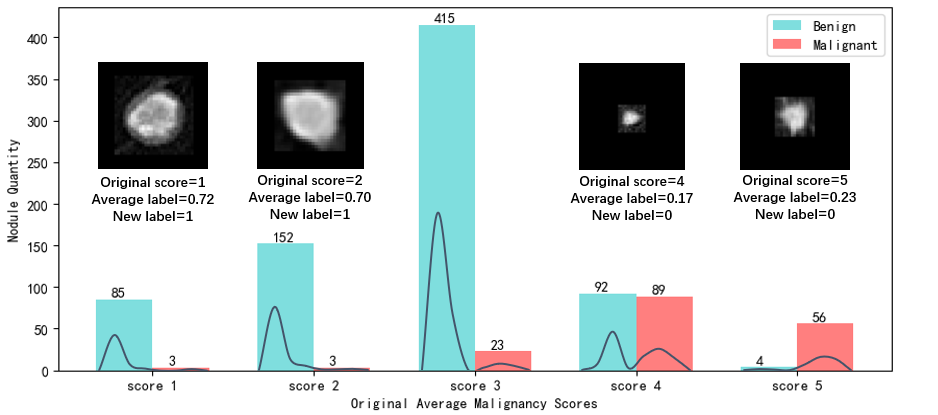}
\caption{Statistical result of LIDC re-labeling nodules (benign or malignant) in terms of original average malignancy scores, where the smooth curve describes the simplified frequency distribution histogram of average label outputs. For each average score of 1, 2, 4, and 5, one nodule re-labeling example with the opposite class (treat score 1 and 2 as benign; treat 4 and 5 as malignant) is provided.}
\label{fig4} 
\end{figure}

We finally re-labeled the LIDC database with the Siamese Network trained using all of SCH-LND. As shown in Fig. \ref{fig4}, our re-labeled results are in broad agreement with the low malignancy score ones. In score 3 (uncertain data), the majority of the nodules are re-labeled to benign class, which explains the better performance when the nodules of score 3 are assigned to benign label in Scenario E (Fig. \ref{fig2}, Case 3). The new labels correct more than half of the original nodule labels with score 4 which could be the main reason leading to the data bias.

\subsection{Discussion}

Re-labeling through metric learning is distinct from the general supervised model in two notable ways. 
First, the input pairs generated by random sampling for metric learning provide a data augmentation effect to overcome overfitting with limited data. 
Second, under-labeled LIDC data take the average labels of top-ranked similarity nodules to increase the confidence of label propagation. These two points may explain why general supervised models (including fine-tuning models) perform worse than metric learning in re-labeling task.
Unfortunately, after re-labeling, the class imbalance problem emerged (748 versus 174), while bringing up new limits in model training performance in the aforementioned experiments.

Moreover, due to the lack of pathological ground truth, the relabel outcomes of this study should always remain suspect until the LIDC clinical information is available. Considering a number of subsequent issues that LIDC may arise, sufficient evidence in this paper explores the motive for us to promote the ongoing collection work of a large pathological-proven nodule database, which is expected to become a powerful open-source database for the international medical imaging and clinical research community.

\section{Conclusion and Future Work}

The LIDC-IDRI database is currently the most popular public database of lung nodules with specific spatial coordinates and experts' annotations. However, because of the absence of clinical information, deep learning models trained based on this database have poor generalization capability in lung cancer prediction and downstream tasks. To challenge the low confidence labels of LIDC, an extra nodule dataset with pathological-proven labels was used to identify the annotation bias problems of LIDC and its label assignment difficulties. With the robust supervision of SCH-LND, we used a metric learning-based approach to re-label LIDC data according to the similar nodule retrieval. The empirical results show that with re-labeled LIDC data, improved performance is achieved along with the maximization of LIDC data utilization and the subsequent class imbalance problem. 
These conclusions provide a guideline for further collection of a large pathological-proven nodule database, which is beneficial to the community.

%
%
%

\begin{thebibliography}{10}
\providecommand{\url}[1]{\texttt{#1}}
\providecommand{\urlprefix}{URL }
\providecommand{\doi}[1]{https://doi.org/#1}

\bibitem{armato2011lung}
Armato~III, S.G., McLennan, G., Bidaut, L., McNitt-Gray, M.F., Meyer, C.R.,
  Reeves, A.P., Zhao, B., Aberle, D.R., Henschke, C.I., Hoffman, E.A., et~al.:
  The lung image database consortium (lidc) and image database resource
  initiative (idri): a completed reference database of lung nodules on ct
  scans. Medical physics  \textbf{38}(2),  915--931 (2011)

\bibitem{bellet2015metric}
Bellet, A., Habrard, A., Sebban, M.: Metric learning. Synthesis Lectures on
  Artificial Intelligence and Machine Learning  \textbf{9}(1),  1--151 (2015)

\bibitem{guo2017learning}
Guo, Q., Feng, W., Zhou, C., Huang, R., Wan, L., Wang, S.: Learning dynamic
  siamese network for visual object tracking. In: Proceedings of the IEEE
  international conference on computer vision. pp. 1763--1771 (2017)

\bibitem{han2015texture}
Han, F., Wang, H., Zhang, G., Han, H., Song, B., Li, L., Moore, W., Lu, H.,
  Zhao, H., Liang, Z.: Texture feature analysis for computer-aided diagnosis on
  pulmonary nodules. Journal of digital imaging  \textbf{28}(1),  99--115
  (2015)

\bibitem{han2013texture}
Han, F., Zhang, G., Wang, H., Song, B., Lu, H., Zhao, D., Zhao, H., Liang, Z.:
  A texture feature analysis for diagnosis of pulmonary nodules using lidc-idri
  database. In: 2013 IEEE International Conference on Medical Imaging Physics
  and Engineering. pp. 14--18. IEEE (2013)

\bibitem{he2016deep}
He, K., Zhang, X., Ren, S., Sun, J.: Deep residual learning for image
  recognition. In: Proceedings of the IEEE conference on computer vision and
  pattern recognition. pp. 770--778 (2016)

\bibitem{hussein2017risk}
Hussein, S., Cao, K., Song, Q., Bagci, U.: Risk stratification of lung nodules
  using 3d cnn-based multi-task learning. In: International conference on
  information processing in medical imaging. pp. 249--260. Springer (2017)

\bibitem{hussein2017tumornet}
Hussein, S., Gillies, R., Cao, K., Song, Q., Bagci, U.: Tumornet: Lung nodule
  characterization using multi-view convolutional neural network with gaussian
  process. In: 2017 IEEE 14th International Symposium on Biomedical Imaging
  (ISBI 2017). pp. 1007--1010. IEEE (2017)

\bibitem{kaya2019deep}
Kaya, M., Bilge, H.{\c{S}}.: Deep metric learning: A survey. Symmetry
  \textbf{11}(9), ~1066 (2019)

\bibitem{kingma2014adam}
Kingma, D.P., Ba, J.: Adam: A method for stochastic optimization. arXiv
  preprint arXiv:1412.6980  (2014)

\bibitem{koch2015siamese}
Koch, G., Zemel, R., Salakhutdinov, R.: Siamese neural networks for one-shot
  image recognition. In: ICML deep learning workshop. vol.~2. Lille (2015)

\bibitem{kramer2011lung}
Kramer, B.S., Berg, C.D., Aberle, D.R., Prorok, P.C.: Lung cancer screening
  with low-dose helical ct: results from the national lung screening trial
  (nlst) (2011)

\bibitem{kumar2017discovery}
Kumar, D., Chung, A.G., Shaifee, M.J., Khalvati, F., Haider, M.A., Wong, A.:
  Discovery radiomics for pathologically-proven computed tomography lung cancer
  prediction. In: International Conference Image Analysis and Recognition. pp.
  54--62. Springer (2017)

\bibitem{liao2021learning}
Liao, Z., Xie, Y., Hu, S., Xia, Y.: Learning from ambiguous labels for lung
  nodule malignancy prediction. arXiv preprint arXiv:2104.11436  (2021)

\bibitem{liu2019multi}
Liu, L., Dou, Q., Chen, H., Qin, J., Heng, P.A.: Multi-task deep model with
  margin ranking loss for lung nodule analysis. IEEE transactions on medical
  imaging  \textbf{39}(3),  718--728 (2019)

\bibitem{mcnitt2007lung}
McNitt-Gray, M.F., Armato~III, S.G., Meyer, C.R., Reeves, A.P., McLennan, G.,
  Pais, R.C., Freymann, J., Brown, M.S., Engelmann, R.M., Bland, P.H., et~al.:
  The lung image database consortium (lidc) data collection process for nodule
  detection and annotation. Academic radiology  \textbf{14}(12),  1464--1474
  (2007)

\bibitem{setio2017validation}
Setio, A.A.A., Traverso, A., De~Bel, T., Berens, M.S., van~den Bogaard, C.,
  Cerello, P., Chen, H., Dou, Q., Fantacci, M.E., Geurts, B., et~al.:
  Validation, comparison, and combination of algorithms for automatic detection
  of pulmonary nodules in computed tomography images: the luna16 challenge.
  Medical image analysis  \textbf{42},  1--13 (2017)

\bibitem{shen2016learning}
Shen, W., Zhou, M., Yang, F., Dong, D., Yang, C., Zang, Y., Tian, J.: Learning
  from experts: Developing transferable deep features for patient-level lung
  cancer prediction. In: International Conference on Medical Image Computing
  and Computer-Assisted Intervention. pp. 124--131. Springer (2016)

\bibitem{shen2015multi}
Shen, W., Zhou, M., Yang, F., Yang, C., Tian, J.: Multi-scale convolutional
  neural networks for lung nodule classification. In: International Conference
  on Information Processing in Medical Imaging. pp. 588--599. Springer (2015)

\bibitem{shen2017multi}
Shen, W., Zhou, M., Yang, F., Yu, D., Dong, D., Yang, C., Zang, Y., Tian, J.:
  Multi-crop convolutional neural networks for lung nodule malignancy
  suspiciousness classification. Pattern Recognition  \textbf{61},  663--673
  (2017)

\bibitem{national2011national}
Team, N.L.S.T.R.: The national lung screening trial: overview and study design.
  Radiology  \textbf{258}(1),  243--253 (2011)

\bibitem{national2011reduced}
Team, N.L.S.T.R.: Reduced lung-cancer mortality with low-dose computed
  tomographic screening. New England Journal of Medicine  \textbf{365}(5),
  395--409 (2011)

\bibitem{wu2019learning}
Wu, B., Sun, X., Hu, L., Wang, Y.: Learning with unsure data for medical image
  diagnosis. In: Proceedings of the IEEE International Conference on Computer
  Vision. pp. 10590--10599 (2019)

\bibitem{wu2018joint}
Wu, B., Zhou, Z., Wang, J., Wang, Y.: Joint learning for pulmonary nodule
  segmentation, attributes and malignancy prediction. In: 2018 IEEE 15th
  International Symposium on Biomedical Imaging (ISBI 2018). pp. 1109--1113.
  IEEE (2018)

\bibitem{wu2018group}
Wu, Y., He, K.: Group normalization. In: Proceedings of the European Conference
  on Computer Vision (ECCV). pp. 3--19 (2018)

\bibitem{xie2017transferable}
Xie, Y., Xia, Y., Zhang, J., Feng, D.D., Fulham, M., Cai, W.: Transferable
  multi-model ensemble for benign-malignant lung nodule classification on chest
  ct. In: International Conference on Medical Image Computing and
  Computer-Assisted Intervention. pp. 656--664. Springer (2017)

\bibitem{xie2018knowledge}
Xie, Y., Xia, Y., Zhang, J., Song, Y., Feng, D., Fulham, M., Cai, W.:
  Knowledge-based collaborative deep learning for benign-malignant lung nodule
  classification on chest ct. IEEE transactions on medical imaging
  \textbf{38}(4),  991--1004 (2018)

\bibitem{zhang2020learning}
Zhang, H., Gu, Y., Qin, Y., Yao, F., Yang, G.Z.: Learning with sure data for
  nodule-level lung cancer prediction. In: International Conference on Medical
  Image Computing and Computer-Assisted Intervention. pp. 570--578. Springer
  (2020)

\bibitem{zhou2019models}
Zhou, Z., Sodha, V., Siddiquee, M.M.R., Feng, R., Tajbakhsh, N., Gotway, M.B.,
  Liang, J.: Models genesis: Generic autodidactic models for 3d medical image
  analysis. In: International Conference on Medical Image Computing and
  Computer-Assisted Intervention. pp. 384--393. Springer (2019)

\end{thebibliography}
%

\end{document}